\documentclass[prl,aps,twocolumn,nofootinbib,preprintnumbers]{revtex4}
\usepackage{graphicx}

\newcommand{\beq}{\begin{equation}}
\newcommand{\eeq}{\end{equation}}
\newcommand{\bea}{\begin{eqnarray}}
\newcommand{\eea}{\end{eqnarray}}
\newcommand{\bear}{\begin{array}}
\newcommand {\eear}{\end{array}}
\newcommand{\bef}{\begin{figure}}
\newcommand {\eef}{\end{figure}}
\newcommand{\bec}{\begin{center}}
\newcommand {\eec}{\end{center}}

\def\GEV#1{10^{#1}{\rm\,GeV}}

\def\lrfp#1#2#3{ \left(\frac{#1}{#2} \right)^{#3}}

\begin{document}
\draft
\tighten
\preprint{TU-935}
\title{\large \bf
Suppressing Isocurvature Perturbations of QCD Axion Dark Matter
}
\author{
    Kwang Sik Jeong\footnote{email: ksjeong@tuhep.phys.tohoku.ac.jp}
    and
    Fuminobu Takahashi\footnote{email: fumi@tuhep.phys.tohoku.ac.jp}}
\affiliation{
 Department of Physics, Tohoku University, Sendai 980-8578, Japan
    }

\vspace{2cm}

\begin{abstract}
We propose a novel mechanism to suppress the isocurvature perturbations of the QCD axion.
The point is that the QCD interactions become strong at an intermediate or high energy scale
in the very early Universe,  if the Higgs field has a sufficiently large expectation value.
The effective QCD scale can be even higher in the presence of extra colored particles.
We show that the QCD axion then becomes so heavy during inflation that its isocurvature
perturbations are significantly suppressed, thereby relaxing the constraint on the inflation scale.
\end{abstract}

\pacs{}
\maketitle

The axion $a$ is a Nambu-Goldstone boson associated with the spontaneous breakdown of the
Peccei-Quinn (PQ) symmetry  introduced to solve the strong CP problem in quantum chromodynamics
(QCD)~\cite{Peccei:1977hh,QCD-axion}. The axion acquires a tiny but non-zero mass from
the QCD instanton effects, and it is stabilized at the CP conserving vacuum.
The dynamical relaxation of the CP phase necessarily induces coherent oscillations of the axion,
which contribute to cold dark matter (CDM) as the axion is stable in a cosmological time scale.

If the axion  is present during inflation, it is subject to quantum fluctuations,
\beq
\delta a \simeq \frac{H_{\rm inf}}{2 \pi},
\eeq
where $H_{\rm inf}$ denotes the Hubble parameter during inflation.
The quantum fluctuations lead to an almost scale-invariant CDM isocurvature density fluctuation,
which leaves a distinctive imprint on the cosmic microwave background (CMB) spectrum.
The observed CMB spectrum can be well fitted by a nearly scale-invariant
adiabatic density perturbation, and a mixture of the isocurvature
perturbation is tightly constrained.
This constraint can be interpreted as an upper bound on the inflation scale in the axion
CDM scenario~\cite{Ade:2013rta},
\beq
H_{\rm inf} < 0.87 \times \GEV{7} \lrfp{f_a}{\GEV{11}}{0.408},
\label{HP}
\eeq
at $95\%~{\rm CL}$,
where $f_a$ is the axion decay constant~\cite{Seckel:1985tj,Lyth:1989pb}.
The constraint becomes much more stringent for a smaller value of $f_a$ due to anharmonic
effects~\cite{Lyth:1991ub,Kobayashi:2013nva}.
Clearly, a chaotic inflation model~\cite{Linde:1983gd,Kawasaki:2000yn,Kallosh:2010ug,Nakayama:2013jka}
is in tension with the axion CDM.

There are several known ways to avoid the constraint.
First, if the PQ symmetry is restored during inflation, the axion does not exist, and so there
is no isocurvature perturbation~\cite{Linde:1990yj,Lyth:1992tx}.
The axion appears when the PQ symmetry is spontaneous broken after inflation.
In such case, however, topological defects such as cosmic strings and domain walls are generated,
and in particular the domain wall number must be unity to avoid the overclosure of the Universe.
Secondly, if the coefficient of the axion kinetic term was larger during inflation
than at present, the size of $\delta a$ can be suppressed.
This can be realized if the radial component
of the PQ scalar has a flat potential and takes a larger value during inflation~\cite{Linde:1990yj,Linde:1991km}.
Alternatively, a similar effect is possible if there is a non-minimal coupling to gravity~\cite{Folkerts:2013tua}.

In this letter we propose another way to relax the isocurvature constraint on the inflation scale
in the axion CDM scenario. The point is that, if the Higgs field has a sufficiently large
expectation value during inflation, the QCD confines at an intermediate or high energy scale
because the quarks decouple at a high energy scale and make the QCD coupling run faster at
lower scales.
Then the axion becomes so heavy that its quantum fluctuations at superhorizon scales
are significantly suppressed, thereby relaxing the constraint on the inflation scale.
The QCD scale can be even higher in the presence of additional colored particles that become
heavy during inflation.

A stronger QCD in the early Universe was considered in Refs.~\cite{Dvali:1995ce,Banks:1996ea}
in order to suppress the cosmological abundance of axions produced by the misalignment mechanism,
which however turned out rather non-trivial to be realized~\cite{Choi:1996fs}.
We will show that the stronger QCD can suppress the axion quantum fluctuations,
but not the abundance in general.

There are a variety of scenarios leading to a large field value of the Higgs fields~\cite{Affleck:1984fy,
Bezrukov:2007ep, Nakayama:2012gh}.
This can naturally be realized in the supersymmetric (SUSY) standard model (SM) which
possesses flat directions involving the Higgs fields $H_u$ and $H_d$.
In this letter we consider a SUSY axion model, but our main idea can be applied
to the non-SUSY case as well as other string theoretic axions in a straightforward way.
In the following we adopt the Planck unit in which $M_{Pl} \simeq 2.4 \times 10^{18}$\,GeV is set to be unity.

We now examine how heavy the axion can be during inflation when the $H_uH_d$
flat direction has a large field value.
In the following we assume that the Higgs fields are neutral under the PQ symmetry.\footnote{
If $H_u$ and $H_d$ are charged under the PQ symmetry, the effective axion decay constant
during inflation is given by $\phi_0$. The isocurvature perturbation constraint can be similarly
relaxed, if one introduces a singlet which gives a mass to additional quarks.
}
The relevant scales are the axion decay constant $f_a$, the effective QCD scale $\Lambda_h$, and
the inflation scale $H_{\rm inf}$.
For simplicity we  assume that the  K\"ahler potential is a generic one such that soft scalar
masses are of order the gravitino mass $m_{3/2}$, and that $f_a$ remains constant
during and after inflation.

Let $\phi^2$ parameterize the $H_uH_d$ flat direction.
It gets a mass correction, $-c H_{\rm inf}^2 |\phi|^2$,
from a Planck-suppressed quartic coupling with the inflaton $X$ in the K\"ahler potential.
The Hubble-induced mass term drives $\phi$ to a large field value for positive $c$
if $H_{\rm inf} > m_{3/2}$.
Then the flat direction is stabilized at $\phi_0 \simeq c^\frac{1}{4}\sqrt{H_{\rm inf} M}$
by the higher order superpotential term,
\beq
\Delta W =
\frac{\phi^4}{4M},
\label{dw}
\eeq
where $M$ is an effective cut-off scale.
We will assume $\phi_0$ to be around or above the GUT scale, $M_{\rm GUT}\simeq 2\times 10^{16}$ GeV,
which can be realized for an appropriate value of $M$. It is also possible to stabilize $\phi$
by a higher order term in the K\"ahler potential.
Note that the mass of the radial component of $\phi$ is of order $H_{\rm inf}$.
The phase component of $\phi$ can have a mass of similar size in the presence of interactions
like $ X|\phi|^2$ or $|X|^2\phi^2$ in the K\"ahler potential.
In this case there are no light degrees of freedom associated with the flat
direction.\footnote{
If the phase of $\phi$ is lighter than $H_{\rm inf}$, it acquires quantum fluctuations, giving
rise to the axion isocurvature perturbations with an amplitude suppressed
by a factor of $f_a/\phi_0$ compared to the conventional scenario.
}

The SUSY SM matter fields including neutrinos acquire large SUSY mass from the Yukawa
couplings with $H_u$ or $H_d$.
Let us first consider a case in which all the quarks are heavier than $\Lambda_h$.
One can then integrate
them out and obtain a pure SUSY SU$(3)_c$ with
the gauge kinetic function \cite{Giudice:1997ni},
\beq
f_h = (\mbox{constant})
- \frac{n}{8\pi^2}\ln S
- \frac{N_f}{8\pi^2}\ln \phi,
\eeq
where $N_f=6$ is the flavor number of the SM quarks, and
$S$ denotes the axion superfield which includes the axion,
\bea
a=\sqrt2\langle |S| \rangle \arg(S) \equiv \frac{f_a}{n}\arg(S).
\eea
The constant $n$ counts the number of PQ charged quarks when the PQ symmetry is linearly
realized.
If $\Lambda_h$ is larger than $H_{\rm inf}$, the gluino condensate is formed during inflation,
giving rise to the non-perturbative superpotential~\cite{Affleck:1983mk},
\bea
W_{\rm np} &=& N_c \Lambda_{0}^3 \,\propto\,  e^{-8\pi^2f_h/N_c}
\nonumber \\
&=& ({\rm constant})\times S^{n/N_c}\phi^{N_f/N_c},
\label{wnp}
\eea
where $N_c=3$, and $\Lambda_0$ denotes the condensation scale which is equal to $\Lambda_h$
in the present case.\footnote{
The Higgs $F$-term $F_\phi/\phi\simeq \sqrt c\, H_{\rm inf}$ induces
a gluino mass via gauge mediation.
Thus $H_{\rm inf}$ should be smaller than $\Lambda_0$ for the gluino
not to decouple at a scale above $\Lambda_0$.
}
The scalar potential of $\phi$ is not affected significantly by $W_{\rm np}$
as long as $H_{\rm inf}\phi^2_0\gg \Lambda_0^3$, which is satisfied for the parameters of our
interest.
Also, $f_a > \Lambda_0$ is needed in order not to destabilize the saxion, as shown below.
Then an axion potential is generated through the non-perturbative
superpotential and SUSY breaking, with the dominant contribution from the Higgs $F$-term:
\bea
\label{axion-potential}
&&
\hspace{-0.8cm}
V = \left|\partial_\phi (\Delta W + W_{\rm np})\right|^2 + \cdots
\nonumber \\
&&
\hspace{-0.4cm}
= 2\sqrt{c}\,N_f
H_{\rm inf} \Lambda_0^3 \cos\left(\frac{a}{N_c f_a} + \theta_0 \right)
+ \cdots,
\eea
for the condensation scale in the range
\bea
H_{\rm inf} < \Lambda_0 < f_a.
\eea
Here $\theta_0$ is generally different from the QCD angle since it receives various contributions
including the phase of $\phi$.
The axion mass thus reads
\beq
\label{axion-mass}
m^2_a = \frac{\hat c H_{\rm inf} \Lambda_0^3}{f^2_a},
\eeq
with $\hat c\equiv 2\sqrt{c}\,N_f/N^2_c$.
Note that a constant term $m_{3/2} M_{Pl}^2$ in the superpotential also
contributes to the axion mass, which is important when $m_{3/2} > H_{\rm inf}$.

On the other hand, if there are $n_f$ quarks lighter than $\Lambda_h$, they form a meson
field stabilized by balancing the Affleck-Dine-Seiberg potential and
the SUSY mass term $m$.
In practice, this is the case for $\Lambda_h \gtrsim \GEV{11}$, where the up and down
quarks are lighter than $\Lambda_h$.
Integrating out the meson fields, one is left with the gluino condensation (\ref{wnp}) with
$\Lambda_{0}^{3 N_c} = \Lambda^{3N_c - n_f}_h {\rm det}(m)$.
Using $\Lambda_0$ defined this way, the axion mass formula (\ref{axion-mass})
holds also in such cases.

The higher QCD scale and the  heavy axion mass during  inflation provide us with a simple way
to suppress the axion CDM isocurvature perturbations.
The axion mass is heavier than the inflation scale, i.e., $m_a > H_{\rm inf}$, if
\beq
\label{Hinf}
H_{\rm inf} <  10^{12}{\rm GeV}\times \hat c
\left(\frac{\Lambda_0}{f_a}\right)^3
\left(\frac{f_a}{10^{12}{\rm GeV}}\right).
\eeq
Then the axion is stabilized at one of the minima of the potential (\ref{axion-potential}),
and no sizable quantum fluctuations are generated at superhorizon scales.
One can see that the isocurvature constraint on the inflation scale can be significantly
relaxed compared to the conventional scenario (\ref{HP}).

So far we have assumed that $f_a$ remains the same during and after inflation.
This is the case if the saxion stabilization is not significantly modified by the strong QCD
interactions, which generically requires $\Lambda_0<f_a$.
To see this, consider a model where the PQ symmetry is broken by two PQ fields $S_1$ and $S_2$
with the superpotential $W=\Sigma(S_1S_2-f^2_0)$ where $\Sigma$ is a PQ singlet.
For $\Lambda_0 < f_0$, the minimum will lie along the $F$-flat direction $S_1S_2=f^2_0$ while
giving $f_a\sim f_0$, if $S_1$ and $S_2$ have soft scalar masses of similar size and couple to
PQ charged quarks.
On the other hand, the saxion potential is significantly affected for $\Lambda_0 > f_0$,
as the non-perturbative superpotential gives the dominant contribution to the saxion potential.
Depending on the axion model, $f_a$ may have a different value from the present one
even if $\Lambda_0 < f_a$.
However the suppression mechanism works as long as the QCD scale is high enough to satisfy
the condition (\ref{Hinf}) for the value of $f_a$ during inflation.

The saxion can be copiously produced by coherent oscillations if its position during inflation
is deviated from the low-energy potential minimum.
If it dominates the energy density of the Universe, axions may be overproduced by the saxion decay.
This can be avoided by e.g. introducing a certain coupling of the saxion to Higgs
multiplet~\cite{Jeong:2012np}.
In fact, the saxion abundance depends on the details of the stabilization of the PQ fields.
In the example considered above, the saxion coherent oscillations after inflation can be suppressed
if there is an approximate symmetry under which $S_1$ and $S_2$ are interchanged.

\begin{figure}[t]
  \begin{center}
  \begin{tabular}{l}
   \includegraphics[width=0.42\textwidth]{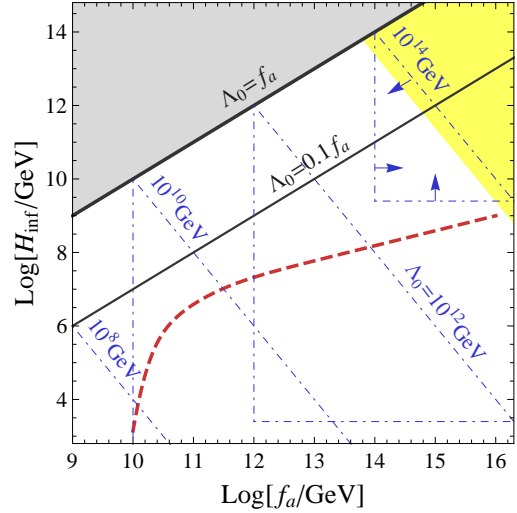}
   \end{tabular}
  \end{center}
  \caption{
  The axion CDM isocurvature perturbation is suppressed by a heavy axion mass
  in the white region. See the text for explanation.
  }
\label{fig:Inflation-scale}
\end{figure}

Fig.~\ref{fig:Inflation-scale} shows which region in the $(f_a,H_{\rm inf})$ plane
is allowed by the constraint $m_a>H_{\rm inf}$, where we take $\hat c=1$ and $\Lambda_0$
as a free parameter.
We will consider an explicit model to realize a large value of $\Lambda_0$ later.
In the upper-left shaded (gray) region, the saxion stabilization is significantly modified
by the QCD interactions as $\Lambda_0$ is larger than $f_a$.
The thick (thin) solid line corresponds to the upper bound on $H_{\rm inf}$ for
$\Lambda_0= f_a$ ($\Lambda_0=0.1 f_a$).
The upper-right yellow region is excluded by the perturbativity limit of the gauge
interactions up to the GUT scale (\ref{H-bound}).
Note that $m_a>H_{\rm inf}$ requires the axion decay constant and the QCD scale during inflation
to satisfy the condition (\ref{Hinf}), which is derived for $H_{\rm inf}>\Lambda^3/\phi^2_0$ and
$H_{\rm inf}<\Lambda_0<f_a$.
Thus, for fixed $\Lambda_0$, the axion CDM isocurvature perturbation is suppressed
inside a right-angled triangle, whose base, height, and hypotenuse
correspond to $\Lambda_0<f_a$, $H_{\rm inf}>\Lambda^3/\phi^2_0$ with $\phi_0\sim M_{\rm GUT}$,
and $m_a>H_{\rm inf}$, respectively.
For instance, the upper-right one with blue dot-dashed sides is obtained for
$\Lambda_0=10^{14}$~GeV.
We also show other cases with $\Lambda_0=10^{8},\,10^{10},\,10^{12}$~GeV, from left to right.
The conventional isocurvature bound (\ref{HP}) is shown by the dashed red line, where
the anharmonic effect~\cite{Kobayashi:2013nva} has been taken into account.

Let us here comment on the axion relic abundance from the misalignment mechanism.
After inflation the axion is located at the minimum of the
potential (\ref{axion-potential}), which is generally different from the CP conserving one.
Hence the axion dark matter abundance today reads \cite{QCD-axion}
\beq
\Omega_a h^2 \simeq 0.2\,\theta^2_i \left(\frac{f_a}{10^{12} {\rm GeV}}\right)^{1.184},
\eeq
with the initial misalignment angle given by
\beq
\theta_i = \theta_0 - \theta_{\rm QCD},
\eeq
where $\theta_{\rm QCD}$ denotes the QCD angle.
For $f_a\gg 10^{12}$ GeV, $|\theta_i|\ll 1$ is needed to avoid overclosure of the Universe.
In other words, the effect of the stronger QCD is to suppress the axion quantum fluctuations,
not the abundance.

The QCD confines at a high energy scale if the Higgs fields are stabilized
at a large field value during inflation, because the quarks obtain large masses.
A similar effect occurs if there are extra colored particles that
become heavier during inflation.
Let us add $N_\Psi$ pairs of $\Psi+\Psi^c$ that belong to
${\bf 5}+\bar{\bf 5}$ of SU$(5)$ but do not carry PQ charges:
\bea
W = \left(M_\Psi + \frac{\phi^2}{M^\prime} \right)\Psi\Psi^c,
\eea
They are fixed at the origin if $M_\Psi+\phi^2/M^\prime$ is larger than $m_{3/2}$ and $H_{\rm inf}$.
For $\phi$ stabilized around the GUT scale, $\Psi+\Psi^c$ become heavy, and therefore the QCD confines
at a larger scale than in the MSSM.
Such effect is maximized when the effective cut-off scale $M^\prime$ is around $M_{\rm GUT}$.
We find that the QCD scale can be as high as,
\beq
\label{QCD-scale}
\Lambda_h \simeq 1.3\times 10^7{\rm GeV} \left(
\frac{M_{\rm GUT}}{M_\Psi}
\right)^{N_\Psi/9},
\eeq
for $\phi_0 \sim M^\prime \sim M_{\rm GUT}$.
The above result shows that $\Lambda_h$ is about $10^7$ GeV when the SM quarks obtain heavy masses
from a large Higgs value around the GUT scale, but it can be raised much higher in the presence
of extra colored particles.
Here we have taken into account the hierarchy in the SM quark masses,
and have used that the universal gauge coupling constant at the GUT scale
is $g^2_{\rm GUT}\simeq 0.5$ in the MSSM.

It is also important to note that $\Psi+\Psi^c$ increase the axion mass since
they decouple while giving a contribution to the coefficient of $\ln \phi$ in $f_h$
in the effective theory.
It is straightforward to see that $\hat c$ is enhanced by a factor $1+2N_\Psi/N_f$,
and therefore the constraint on $H_{\rm inf}$ is further relaxed.

On the other hand, in the axion models where the PQ symmetry is linearly realized,
there should exist PQ charged quarks in order for $S$ to couple to the QCD anomaly.
Let us introduce $N_\Phi$ pairs of $\Phi+\Phi^c$ transforming as ${\bf 5}+\bar{\bf 5}$
under SU$(5)$, and consider the superpotential term $S\Phi\Phi^c$.
Then $\Phi+\Phi^c$ obtain large masses $M_\Phi\sim f_a$.
The gauge coupling unification is maintained in the presence of additional
${\bf 5}+\bar{\bf 5}$ matter fields, but the gauge coupling at the GUT scale is larger
than the MSSM value.
In order for the gauge interactions to remain perturbative up to $M_{\rm GUT}$,
we need
\bea
\frac{1}{g^2_{\rm GUT}} \simeq
2 - \sum_{i=\Phi,\Psi}
\frac{N_i}{8\pi^2}\ln\left(\frac{M_{\rm GUT}}{M_i}\right)
> \frac{1}{4\pi}.
\eea
Note that the PQ charged quarks play no role in raising the QCD scale because their masses
remain the same during and after inflation.
The above requirement is combined with the relation (\ref{QCD-scale}) to put the upper bound on
the QCD scale,
\beq
\label{H-bound}
\Lambda_h < 2.7\times 10^{14}{\rm GeV}
\left(\frac{M_\Phi}{M_{\rm GUT}}\right)^{N_\Phi/9},
\eeq
which is insensitive to $M_\Psi$.
Thus the QCD scale can be high enough to achieve $m_a>H_{\rm inf}$ in the early Universe.
For instance, for $M_\Phi\geq 10^{9}$ GeV and $N_\Psi=1$, $\Lambda_0$ can be as
high as about $10^{13}$ GeV.
We note that $\Lambda_0$ is slightly smaller than $\Lambda_h$ if
$\Lambda_h\gtrsim 10^{-5}M_{\rm GUT}$, due to the hierarchy of the SM quark masses.

\begin{figure}[t]
  \begin{center}
  \begin{tabular}{l}
   \includegraphics[width=0.42\textwidth]{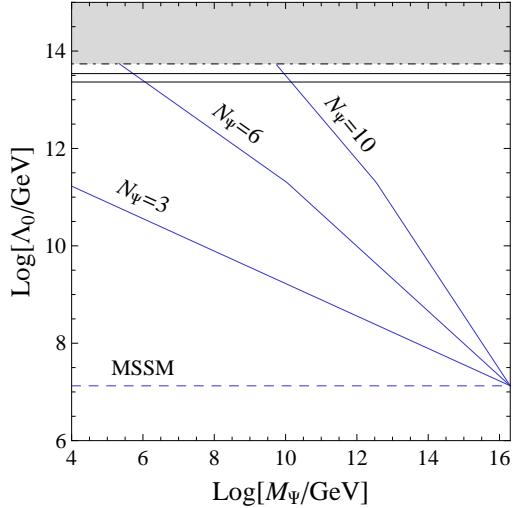}
   \end{tabular}
  \end{center}
  \caption{
  Condensate scale $\Lambda_0$ for $\phi$ fixed at $M_{\rm GUT}$ in the presence
  of extra ${\bf 5}+\bar{\bf 5}$ matter fields. See the text for explanation.
  }
\label{fig:QCD-scale}
\end{figure}

Fig.~\ref{fig:QCD-scale} shows how much $\Lambda_0$ can be raised for $\phi_0=M_{\rm GUT}$
during inflation, under the requirement of the perturbativity of gauge couplings up to
the GUT scale.
The solid (blue) lines represent $\Lambda_0$ for $N_\Psi=3, 6, 10$  from bottom to top,
while the dashed one is for the MSSM without extra matter fields.
In the shaded region above the dot-dashed black line,
the perturbativity of gauge couplings breaks down below $M_{\rm GUT}$.
This bound becomes severer in the presence of PQ charged fields $\Phi+\Phi^c$:
the upper (lower) solid-black line is the bound for the case
with $N_\Phi=1$ and $M_\Phi=10^{14}$ GeV ($M_\Phi=10^{12}$ GeV).

The axion mass depends on $c$, the coupling of $\phi$ with the inflaton.
If $c \gtrsim {\cal O}(10)$, the constraint on the Hubble parameter can be
relaxed further, making the chaotic inflation consistent with the axion dark matter.

So far we have focused on the case of $m_a > H_{\rm inf}$.
If $m_a < H_{\rm inf}$, the axion acquires quantum fluctuations.
It is, however, possible that the axion remains heavy for a while even
after inflation, and starts to oscillate and decays into photons.
Then the isocurvature constraint can be avoided. To this end, one needs to
delay the commencement of oscillation of $\phi$ as well as thermalization of the SM particles.
The former requires a rather flat potential of $\phi$ at large field values, while the latter
is possible if the inflaton decays into hidden particles, not into the SM particles.

Lastly let us mention the cosmological evolution after inflation. The Universe will
be reheated by the decay of either the inflaton or $\phi$.
In the latter case, the preheating as well as subsequent dissipation effects will
be important if $\phi$ passes through the origin during
oscillations~\cite{GarciaBellido:2008ab,Mukaida:2012bz}.

In this letter we have proposed a novel mechanism to suppress the axion CDM
isocurvature perturbations. The point is that, if QCD becomes strong at an intermediate
or high scale during inflation, the axion can be so heavy that no sizable quantum fluctuations
at superhorizon scales are produced.

\section*{Acknowledgment}
This work was supported by Grant-in-Aid for Scientific Research (C) (No. 23540283) [KSJ],
Scientific Research on Innovative Areas (No.24111702 [FT], No. 21111006 [FT],
and No.23104008 [KSJ and FT]), Scientific Research (A) (No. 22244030 and No.21244033) [FT],
and JSPS Grant-in-Aid for Young Scientists (B) (No. 24740135) [FT].

\end{document}